\documentclass[%
 reprint,
 superscriptaddress,
 amsmath,amssymb,
 aps,
 pra,
]{revtex4-2}

\usepackage{easyReview}
\usepackage{graphicx}
\usepackage{dcolumn}
\usepackage{bm}
\usepackage[utf8]{inputenc}
\usepackage{textcomp}
\DeclareUnicodeCharacter{2264}{\ensuremath{\leq}}
\DeclareUnicodeCharacter{2212}{\ensuremath{-}}
\DeclareUnicodeCharacter{2248}{\ensuremath{\approx}}
\usepackage[version=4]{mhchem}
\usepackage{enumitem}
\usepackage{csquotes}
\usepackage{nicefrac}
\usepackage{mathtools}
\usepackage{xcolor}
\usepackage{soul}
\usepackage[normalem]{ulem}
\usepackage{siunitx}
\usepackage{url}
\usepackage[colorlinks=true, 
            linkcolor=black, 
            urlcolor=blue, 
            citecolor=black]{hyperref}
\usepackage{cleveref}
\crefname{figure}{Fig.}{Figs.}
\Crefname{figure}{Figure}{Figures}

\crefname{equation}{Eq.}{Eqs.}
\Crefname{equation}{Equation}{Equations}

\crefname{section}{Sec.}{Secs.}
\Crefname{section}{Section}{Sections}

\crefname{table}{Table}{Tables}
\Crefname{table}{Table}{Tables}

 \renewcommand{\selectlanguage}[1]{} 

\renewcommand{\arraystretch}{1.2}

\DeclareSIUnit\permille{\text{\textperthousand}}

\newcommand{\hetwo}{\ce{^{4}He^{2+}}}

\newcommand{\cfour}{\ce{^{12}C^{4+}}}
\newcommand{\csix}{\ce{^{12}C^{6+}}}

\DeclareSIUnit\u{u}
\newcommand{\ebgmedaustrongmbh}{MedAustron, Marie-Curie Straße 5, 2700 Wiener Neustadt, Austria}
\newcommand{\ati}{Atominstitut, TU Wien, Stadionallee 2, 1020 Vienna, Austria}
\newcommand{\meduni}{Medical University of Vienna, Spitalgasse 23, 1090 Vienna, Austria}
\newcommand{\gsi}{GSI Helmholtzzentrum für Schwerionenforschung GmbH, Planckstraße 1, 64278 Darmstadt, Germany}

\begin{document}

\preprint{APS/123-QED}

\title{{Noninvasive ion fraction quantification of dual-species beams in synchrotrons}}

\author{Elisabeth Renner}
    \affiliation{\ati}
    \email{Contact author: elisabeth.renner@tuwien.ac.at}

\author{Matthias Kausel}
    \affiliation{\ebgmedaustrongmbh}
    \affiliation{\ati}

\author{David Ondreka}
    \affiliation{\gsi}

\author{Hermann Fuchs}
    \affiliation{\meduni}

\author{Katrin Holzfeind}
    \affiliation{\ati}

\author{Nana Okropiridze}
    \affiliation{\ati}


\begin{abstract}
The ion composition of dual-species beams in synchrotrons is typically inferred from invasive measurements performed after beam extraction.
This paper introduces a complementary noninvasive method to determine the ion composition of such beams directly inside the synchrotron. The approach is applicable to low- and medium-energy ion synchrotrons and to small relative mass-to-charge ratio offsets, typically at the $10^{-4}$ level. The method exploits dispersive orbit offsets between the two species and corresponding frequency corrections applied by the synchrotron RF radial regulation loop. This capability is of particular interest for ongoing research on online monitoring in carbon ion beam therapy using mixed $^4$He$^{2+}$ and $^{12}$C$^{6+}$ beams, which feature a relative mass-to-charge ratio offset of 0.065\,\%.

The proposed method is analytically derived and tested with particle tracking simulations using Xsuite. Its applicability under realistic experimental conditions is demonstrated at the MedAustron facility using mixed $^4$He$^{2+}$ and $^{12}$C$^{6+}$ beams. The results show good agreement with independent post-extraction measurements.
\end{abstract}

\maketitle

\section{Introduction}

\begin{figure}
    \includegraphics[trim={0cm 0cm 0cm 0cm},
    clip,    
    width=0.6\linewidth]{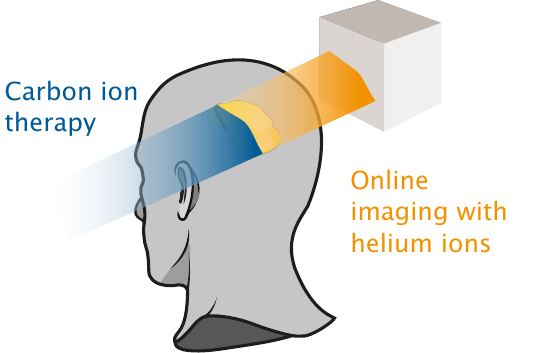}
    \caption{Schematic mixed-beam irradiation. The carbon beam is used for tumor treatment while the residual energy of the helium ions is measured downstream of the patient for diagnostic purposes.}\label{fig:schematicMixedBeamIrrad}
\end{figure}
In recent years, mixed helium and carbon ion irradiation has been proposed as a new concept for online range monitoring in carbon ion therapy~\cite{mazzucconi_mixed_2018,graeff_helium_2018}.
The aim of such monitoring is to detect range errors, for example due to anatomical changes or uncertainties associated with CT-to-stopping-power conversion. 

\hetwo{} and \csix{} have a similar mass-to-charge ratio $m/q$, with a relative difference of
\begin{equation}\label{eq:eq:rel-m-q}
    \dfrac{\Delta \left(m/q\right)}{\left(m/q\right)} = \dfrac{\left(m/q\right)_\mathrm{He}-\left(m/q\right)_\mathrm{C}}{\left(m/q\right)_\mathrm{C}} = 6.5 \times 10^{-4}~. 
\end{equation}
This allows both ion species to be simultaneously  delivered to the patient at a similar energy-per-mass, at which \hetwo{} has approximately three times the range in matter of \csix{}. 
As a result, the tumor can be treated with carbon ions while range errors are inferred from the residual energy of the helium ions traversing the patient (\cref{fig:schematicMixedBeamIrrad}).

Since 2023, both at GSI~\cite{galonska_first_2024} and MedAustron~\cite{kausel_prab,renner:ipac2026-tuo8t01}, mixed helium and carbon ion beams have been successfully generated, accelerated, and extracted to experimental irradiation rooms for interdisciplinary research in accelerator physics, detector development, and medical physics. Due to differing accelerator capabilities, GSI currently generates the mixed beam directly in a single ion source~\cite{galonska_first_2024}, while at MedAustron the two ion species are generated in separate ion sources and mixed by means of a sequential multiturn injection into the synchrotron~\cite{kausel_prab}. 

Currently, these beams are provided to research users on an experimental basis, with helium fractions
\begin{equation}
    r = \dfrac{N_\mathrm{He}}{N_\mathrm{He}+N_\mathrm{C}}
\end{equation}
typically varying within ${\num{0.1}<r<\num{0.5}}$, depending on the specific user requirements. $N_\mathrm{He}$ and $N_\mathrm{C}$ are the  intensities per species. Studies towards future clinical implementation currently consider helium fractions of ${r \approx \qtyrange{0.1}{0.2}{ }}$, as this range is expected to provide a helium signal distinguishable from the carbon-ion fragments, while the dose deposited by the helium ions amounts to less than \SI{1}{\percent} of the total dose~\cite{mazzucconi_mixed_2018, hardt2025}. However, final clinical target values and tolerances for the helium fraction have not yet been established.

If the beam is generated from a single ion source (e.g., using a helium-methane mixture as source gas), the helium fraction can be controlled by adjusting the helium gas flow to the source plasma~\cite{galonska_first_2024}. In the case of mixed-beam generation in of a sequential multiturn injection,  as done at MedAustron~\cite{kausel_prab}, it is possible to tailor the helium fraction using the injection settings (\cref{sec:mixedbeam_gen_medaustron}). However, particularly in the latter approach, the helium fraction can exhibit noticeable shot-to-shot fluctuations. 

At present, the ion composition cannot be determined directly within the accelerator complex and is instead inferred from measurements of the extracted beam in the irradiation room. Such measurements require the installation of a dedicated diagnostic setup, which complicates beam commissioning tasks. Further, in most cases the setup cannot be installed simultaneously with experiments, which means that no cycle-resolved information on the delivered mixing ratio is available.

To address this gap, this work proposes a new noninvasive method for estimating the ion composition prior to extraction in synchrotrons. The method relies on composition-dependent feedback from the radial regulation loops of the synchrotron radio frequency (RF) system.
Due to the small but measurable difference in beam rigidity between \hetwo{} and \csix{}
(\cref{sec:rfacc_mixed}), the two ion species follow different dispersive orbits. In synchrotrons without radial RF regulation, information on the ion composition can be obtained from the measured orbit offset in dispersive regions. In synchrotrons with radial RF regulation it can be retrieved from the applied frequency corrections. While the present work focuses on the latter case, the methodology can in principle also be adapted to synchrotrons without radial regulation loops.

\Cref{sec:Background} reviews the status of mixed-beam delivery at MedAustron as well as the RF acceleration of such beams. \Cref{sec:model} presents the proposed method for determining the ion fraction in the synchrotron.
The experimental demonstration is described in \cref{sec:experiment}, followed by discussion  and concluding remarks in \cref{sec:discussionoutlook,sec:conclusion}.


\section{Background}\label{sec:Background}

\subsection{Mixed-ion beam delivery at MedAustron}\label{sec:mixedbeam_gen_medaustron}

At the MedAustron center for ion beam therapy and research,
which is based on the Proton-Ion Medical Machine Study (PIMMS)~\cite{badano_proton-ion_2000, bryant_proton-ion_2000}, mixed helium-carbon ion beams are generated and delivered to the irradiation room for non-clinical research studies using a sequential multiturn injection scheme~\cite{kausel_prab}. In nominal single-species operation, helium and carbon ion beams are generated in two separate ion sources as ${\approx 8}$\,keV/u \hetwo{} and \cfour{} beams, which are subsequently pre-accelerated through a radio frequency quadrupole and linear accelerator to $\approx\SI{7}{\mega\electronvolt\per\u}$. The beams then pass a stripping foil. The fully stripped \hetwo{} and \csix{} ions are injected into the synchrotron  via a conventional multiturn injection.

In mixed-species operation, the beam is generated by successively injecting the \hetwo{} and \csix{} ions from the two distinct ion sources into the synchrotron. First, helium is injected and stored at flat bottom. After this, the injector is reconfigured and the carbon beam is injected using a second multiturn injection. A substantial fraction of the already circulating helium ion beam is lost at the injection septum when ramping the injection bump for the carbon injection. As a result, the two ion species  end up occupying different regions in horizontal phase space. One way to tailor the ion fraction is to adjust the amplitude of the second injection bump~\cite{kausel_prab}. 

Following sequential injection, both ion species are captured together and accelerated  to extraction energies between 120\,MeV/u and \SI{402.8}{\mega\electronvolt\per\u}. As in single-species operation, the synchrotron RF system~\cite{schmitzer:ipac16-mopoy001} employs radial and phase regulation loops. The system operates with a loop cycle time of 10\,{\textmu}s. The two ion species are subsequently extracted together using third-order resonance slow extraction. To date, the clinically employed betatron core-driven slow extraction is not compatible with mixed-beam operation. Therefore the mixed-beam  slow extraction has so far  been performed~\cite{renner:ipac2026-thp4091} with either RF knockout or phase displacement acceleration as driving mechanism, profiting from recent developments on these alternative slow extraction mechanisms at MedAustron~\cite{DeFranco:IPAC2018-MOPML025, kuhteubl_slow_2024, renner_investigating_2024}.

\subsection{RF acceleration of mixed-species ion beams}\label{sec:rfacc_mixed}

When accelerating mixed-species beams with slightly different $m/q$ in synchrotrons the RF synchronism condition must be simultaneously fulfilled for all ion species despite the slightly different closed-orbit lengths. As a consequence, the secondary ion species attains a slight offset in momentum-per-mass from the reference species, as reviewed below.

The synchronous particle of the reference species (species~I) is described using the rest mass $m_\mathrm{I}$, charge $q_\mathrm{I}$, velocity $v_\mathrm{I} = \beta_\mathrm{I} \,c$, with the speed of light $c$, and the relativistic factors  $\beta_\mathrm{I}$ and $\gamma_\mathrm{I}$. The synchronous particle of the secondary ion species (species~II) is described using $m_\mathrm{II}$, $q_\mathrm{II}$, $\beta_\mathrm{II}$ and $\gamma_\mathrm{II}$. Note that the synchronous particle of species~I may be off-momentum with respect to the on-momentum reference particle, which is denoted by the subscript~``0''. The label~``sp''  is used to indicate offsets between the two ion species.  

The relative offset in magnetic rigidity {$B \rho = \frac{m \gamma \beta c}{q}$} of species~II from species~I,
\begin{equation}\label{eq:dpq}
    \begin{aligned}
        \dfrac{\Delta(B\rho)}{\left(B\rho\right)_\mathrm{I}} &= \dfrac{\Delta \left(p/q\right)}{\left(p/q\right)_\mathrm{I}} = \dfrac{m_\mathrm{II}/q_\mathrm{II}}{m_\mathrm{I}/q_\mathrm{I}} \cdot \dfrac{\beta_\mathrm{II} \gamma_\mathrm{II}}{\beta_\mathrm{I} \gamma_\mathrm{I}}-1 \\[2mm]
        &=\delta_\mathrm{eff,sp}~, 
    \end{aligned}
\end{equation}
depends on offsets in both, $m/q$ and ${p/m = \beta \gamma c}$. Following the notation of Hermes et al.~\cite{hermes_heavy-ion_2016}, it is also referred to as \textit{effective relative momentum offset} $\delta_\mathrm{eff}$.    

Similar to the single-species case, a deviation in $p/m$, i.e. in $\beta\gamma$, will be parametrized by
\begin{equation}\label{eq:delta_from_betagamma}
    \begin{aligned}
        \delta_\mathrm{sp} & = \dfrac{\Delta \left(p/m\right)}{\left(p/m\right)_\mathrm{I}}= \dfrac{\Delta \left(\beta \gamma\right)}{\left(\beta \gamma\right)_\mathrm{I}}= \dfrac{\beta_\mathrm{II} \gamma_\mathrm{II}}{\beta_\mathrm{I} \gamma_\mathrm{I}} -1~.\\
    \end{aligned} 
\end{equation}
The offset in $m/q$ is commonly described by the factor 
\begin{equation}\label{eq:chi}
    \chi_\mathrm{sp} = \dfrac{q_\mathrm{II}/m_\mathrm{II}}{q_\mathrm{I}/m_\mathrm{I}}~.
\end{equation}
Together, \cref{eq:delta_from_betagamma,eq:chi} are used to rewrite the relative rigidity offset between species~I and~II, \cref{eq:dpq}, as
\begin{equation}\label{eq:delta_eff}
    \delta_\mathrm{eff,sp} = \dfrac{1+\delta_\mathrm{sp}}{\chi_\mathrm{sp}} -1~.
\end{equation}
The formalism presented in \cref{sec:model} is based on linearized models valid for small mass-to-charge ratio offsets $\left|\chi_\mathrm{sp} -1\right|\ll 1$. Treating derivations in this work consistently to first order, \cref{eq:delta_eff} can be expressed as
\begin{equation}\label{eq:delta_eff_lin}
    \delta_\mathrm{eff,sp} \approx \dfrac{1}{\chi_\mathrm{sp}}  -1 + \delta_\mathrm{sp}~.
\end{equation}
Due to this rigidity offset, the secondary ion species follows a different dispersive orbit than the reference ion species, with the modified average bending radius
\begin{equation}\label{eq:RR0}
    R_\mathrm{II} = R_\mathrm{I} \, \left(1 + \alpha_{\mathrm{c}} \,\delta_\mathrm{eff,sp}\right) + \mathcal{O}\left(\delta_\mathrm{eff,sp}^2\right)~,
\end{equation}
where $\alpha_{\mathrm{c}}$ is the linear momentum compaction factor. In the case of off-momentum operation $\alpha_{\mathrm{c}}$  is evaluated at the momentum offset of the reference species~I. 

When the RF system is turned on, both synchronous particles must have the same revolution frequency~$f_\mathrm{rev}$, despite their different closed-orbit lengths. This results in the two ion species acquiring slightly different $\beta$, i.e.\
\begin{equation}\label{eq:dff}
    \begin{aligned}
        \dfrac{\Delta f_\mathrm{rev}}{f_\mathrm{rev,I}} &= 0  = \dfrac{\beta_\mathrm{II}}{\beta_\mathrm{I}} \dfrac{R_\mathrm{I}}{R_\mathrm{II}} -1\\
        &\Rightarrow \dfrac{\beta_\mathrm{II}}{\beta_\mathrm{I}} = \dfrac{R_\mathrm{II}}{R_\mathrm{I}}~.
    \end{aligned}
\end{equation}
Inserting \cref{eq:RR0}, with \cref{eq:delta_eff_lin}, and 
\begin{equation}\label{eq:beta_expansion}
    \dfrac{\beta_\mathrm{II}}{\beta_\mathrm{I}} \approx 1 + \dfrac{1}{\gamma_\mathrm{I}^2} \delta_\mathrm{sp} + \mathcal{O}\left(\delta_\mathrm{sp}^2\right)~,
\end{equation}
into \cref{eq:dff} yields an expression for the resulting relative offset in $p/m=\beta \gamma c$ of species~II from species~I,
\begin{equation}\label{eq:hat_delta}
\begin{aligned}
    {\delta}_\mathrm{sp} &= \dfrac{(\beta\gamma)_\mathrm{II}-(\beta\gamma)_\mathrm{I}}{(\beta\gamma)_\mathrm{I}}\\[2mm]
    &= \dfrac{\gamma_\mathrm{I}^2}{\gamma_\mathrm{tr}^2 -  \gamma_\mathrm{I}^2} \, \left(\dfrac{1}{\chi_\mathrm{sp}}-1\right)~,
\end{aligned}
\end{equation}
with $\gamma_\mathrm{tr}=1/\sqrt{\alpha_{\mathrm{c}}}$ being the transition energy.

Combining~\cref{eq:hat_delta} and~\cref{eq:delta_eff_lin}, we obtain an explicit expression for the effective momentum deviation, valid when the secondary ion species circulates at the same frequency as the reference species,
\begin{equation}
     \delta_\mathrm{eff,sp} = 
     \dfrac{1}{1 - \gamma_\mathrm{I}^2/\gamma_\mathrm{tr}^2} \, \left(\dfrac{1}{\chi_\mathrm{sp}}-1\right)~.
\end{equation}

Alternatively, \cref{eq:hat_delta} can also be obtained by logarithmic differentiation, i.e. 
\begin{equation}
    \dfrac{\mathrm{d}\left(p/q\right)}{\left(p/q\right)}=\dfrac{\mathrm{d}\left(m/q\right)}{\left(m/q\right)}+
    \dfrac{\mathrm{d}\left(\beta\gamma\right)}{\left(\beta\gamma\right)}~,
\end{equation} 
as presented for example in references~\cite{bryant_principles_1993, angert_accelerating_1988}. 

As an example, we consider a mixed \hetwo{} and \csix{} beam in a PIMMS-like synchrotron with ${\alpha_c \approx 0.25}$, operated below transition. \csix{} is defined as the reference species~I. \hetwo, species~II, features  ${1 - \chi_\mathrm{sp}= 6.5 \times 10^{-4}}$. It is therefore slightly more rigid than \csix{} and follows an orbit with increased path length. This means that, below transition, \hetwo{} must be accelerated to slightly higher velocities, $\beta_\mathrm{He} > \beta_\mathrm{C}$, to ensure synchronism with the RF wave. The introduced relative offset in $\beta\gamma$  for \hetwo{} in such a synchrotron, \cref{eq:hat_delta},  is between ${\delta}_\mathrm{sp}\approx \SI{0.3}{\permille}$ and $\SI{0.7}{\permille}$ for clinical extraction energies of ${E_\mathrm{kin}=\qtyrange{120}{400}{\mega\electronvolt\per\u}}$. This offset is comparable in magnitude to the operational rms momentum spreads $\delta_\mathrm{rms}$ at flat top, prior to an optional phase jump~\cite{holzfeind2024implementation}. The related total effective rigidity offset, \cref{eq:delta_eff_lin}, ranges from \SI{0.95}{\permille} to \SI{1.35}{\permille} for the stated extraction energies.

Above transition, \cref{eq:hat_delta}  changes sign which reduces the effective rigidity offset, \cref{eq:delta_eff_lin}. For ${\gamma_\mathrm{I} \gg \gamma_\mathrm{tr}}$, $\delta_\mathrm{sp} \rightarrow \left(1-1/\chi_\mathrm{sp}\right)$, the two contributions in  \cref{eq:delta_eff_lin} compensate and $\delta_\mathrm{eff,sp}$ vanishes. Both ion species move on a similar orbit. 
However, when approaching transition energy the orbit separation increases. Whether both ion species in a mixed-species beam can be maintained during transition crossing would therefore depend on the details of the transition crossing scheme and the value of $\chi_\mathrm{sp}$. This effect was made use of in the late 1980s in the CERN PS to clean a $\ce{S}^{16+}$ beam from an $\ce{O}^{8+}$ contamination~\cite{angert_accelerating_1988,hancock_transition_1990}.


\section{Ion fraction quantification}\label{sec:model}

\begin{figure*}
    \includegraphics[width=\textwidth]{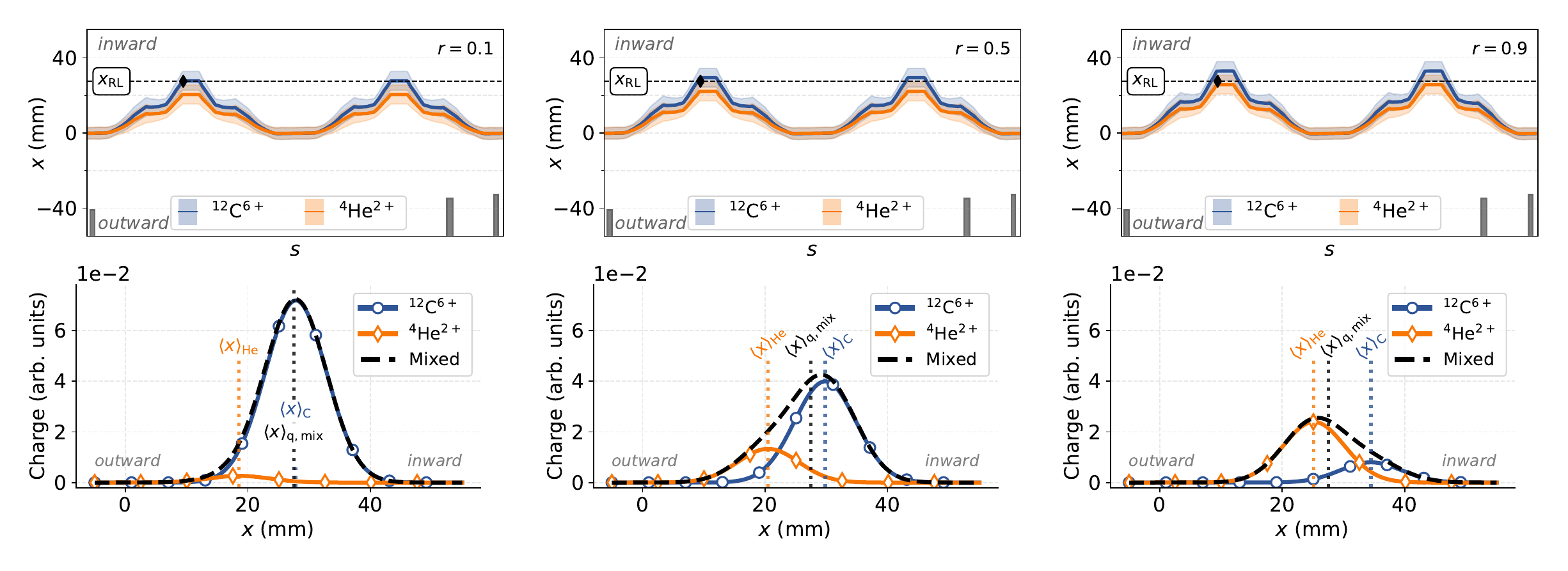}
    \caption{Mixed-beam charge distributions for different helium fractions $r$. The radial loop regulates the horizontal center of gravity of the mixed charge distribution, $\left\langle x\right\rangle_\mathrm{q,mix}$, to $x_\mathrm{RL}=\qty{27.5}{\milli\metre}$.
    Top: 1$\sigma_\mathrm{rms}$ beam envelopes along a PIMMS lattice. Bottom: Horizontal charge distributions at the radial loop pick-up with $D_x=\qty{-8.3}{\metre}$.}
    \label{fig:schematic_deltaR}
\end{figure*}

This section presents a model that exploits the described difference in beam rigidity between two ion species to noninvasively determine the mixing ratio. The model is applicable to low- and medium-energy ion synchrotrons, ion mixtures with small mass-to-charge offsets, ${\left|\chi_\mathrm{sp}-1\right|=\mathcal{O}(10^{-4})}$, and is formulated under the assumption that the horizontal dispersion is approximately constant for the respective rigidity offsets. Furthermore, it requires that the beam position monitors exhibit an approximately linear response across the horizontal range occupied by both ion species, allowing their signals to be superimposed. This condition is met by shoe-box type pick-ups~\cite{Strehl:2006sho}.

A proposal conceptually related to the approach pursued here was put forward by S.~van~der~Meer in 1978~\cite{vanderMeer:119411}.
He suggested determining the mass difference between $p$ and $\bar{p}$ in the CERN SPS by resolving their orbit separation in a dispersive region when operating close to transition energy. However, this approach was ultimately abandoned, as subsequent analyses indicated that for that specific use case alternative methods could achieve higher accuracy~\citep[Chp. 7.7]{bryant_principles_1993}.

\subsection{Dispersive distribution of mixed-species beams}

For a dual-species beam with $N_\mathrm{I}$ ions of species~I with charge $Q_\mathrm{I} = q_\mathrm{I}\, e$ (charge state $q$, elementary charge $e$) and $N_\mathrm{II}$ ions of species~II with charge state $q_\mathrm{II}$, it is useful to distinguish between the total number of ions
\begin{equation}
    N_\mathrm{mix} = N_\mathrm{I} + N_\mathrm{II}~,
\end{equation}
and the total number of charges
\begin{equation}
    N_{\mathrm{q,mix}} = N_\mathrm{I}\, q_\mathrm{I} + N_\mathrm{II}\, q_\mathrm{II}~.
\end{equation}
Using the previously introduced definition for the ion fraction 
\begin{equation}
    r = \dfrac{N_\mathrm{II}}{N_\mathrm{II}+N_\mathrm{I}} =\dfrac{N_\mathrm{II}}{N_\mathrm{mix}}~,
\end{equation}
$N_{\mathrm{q,mix}}$ can be expressed by
\begin{equation}
\begin{aligned}
    N_\mathrm{q,mix} &= N_\mathrm{I} \, q_\mathrm{I} +N_\mathrm{II}  \, q_\mathrm{II}\\[2mm]
    &=  
    \left(1-r\right)\, N_\mathrm{mix}\,q_\mathrm{I} +r\,N_\mathrm{mix} \, q_\mathrm{II}~. \\
\end{aligned}
\end{equation}

At a location with non-zero horizontal dispersion ${D_x\neq 0}$, the center of gravity of species~II features a horizontal offset from the center of gravity of species~I of 
\begin{equation}\label{eq:dx_he_c}
    \left\langle x\right\rangle_\mathrm{II} - \left\langle x\right\rangle_\mathrm{I} = D_x \, \delta_\mathrm{eff,sp}, 
\end{equation}
with $\delta_\mathrm{eff,sp}$ obtained from \cref{eq:delta_eff_lin,eq:hat_delta}.
In the case of off-momentum operation, with the reference ion species featuring a relative momentum offset $\delta_\mathrm{I}$ (as common in a PIMMS-like synchrotron), $D_x$ is the horizontal dispersion evaluated at the momentum offset of the reference ion species, i.e, $D_x=D_x(\delta_\mathrm{I})$.

Consequently, the two displaced charge distributions shift the center of gravity position of the entire charge distribution to
\begin{equation}\label{eq:avg_x_q_mix}
    \begin{aligned}
    \left\langle x\right\rangle_\mathrm{q,mix} &= \dfrac{N_\mathrm{II} \, q_\mathrm{II} \, \left\langle x\right\rangle_\mathrm{II} + N_\mathrm{I} \, q_\mathrm{I} \, \left\langle x\right\rangle_\mathrm{I}}{N_\mathrm{q,mix}} \\[2mm]
     &= A_\mathrm{r} \, D_x \, \delta_\mathrm{eff,sp}+ \left\langle x\right\rangle_\mathrm{I}~,\\
    \end{aligned}
\end{equation}
with the charge-weighted ion fraction 
\begin{equation}
    A_\mathrm{r} = \dfrac{q_\mathrm{II}\, r }{q_\mathrm{II} \, r + q_\mathrm{I} \, (1-r)}.
\end{equation}

To provide an example, \cref{fig:schematic_deltaR} illustrates the simulated charge distributions for a mixed \SI{262.3}{\mega\electronvolt\per\u} \hetwo{} and \csix{} beam with different ion fractions $r$ in a PIMMS~\cite{badano_proton-ion_2000, bryant_proton-ion_2000} lattice. \csix{} is considered the reference ion species. As in single-species operation, the modeled synchrotron is operated off-momentum, with a radial loop correcting the RF frequency so that the measured center of gravity of the total charge distribution, $\left\langle x\right\rangle_\mathrm{q, mix}$, features the requested offset $x_\mathrm{RL}$ at the radial loop pick-up, in this example $x_{\mathrm{RL}}=\SI{27.5}{\milli\meter}$. The two distributions are separated by the offset given in \cref{eq:dx_he_c}. 
In the convention used throughout this work, the beam moves counterclockwise and hence ${x>0}$ and ${x<0}$ denote horizontal offsets toward and away from the center of the synchrotron, respectively, which is indicated in \cref{fig:schematic_deltaR} as \textit{inward} and \textit{outward}. \hetwo{}, being more rigid, with $\delta_\mathrm{eff,sp}>0$ moves on a closed orbit shifted towards the outside of the synchrotron.

The top row displays the horizontal $1\sigma_{\mathrm{rms}}$ beam envelopes of \hetwo{} (orange) and \csix{} (blue). The black dashed line marks the radial loop setpoint. The diamond marker indicates the dispersive location (${D_x=\SI{-8.3}{\meter}}$) at which a representative pick-up is installed in this model. 
The envelope is computed as $\sigma_{\mathrm{rms}} = \sqrt{\beta_x \epsilon_{x, \mathrm{n,rms}}/(\gamma\beta) + D_x^2 \delta_{\mathrm{rms}}^2}$,
where $\beta_x$ denotes the Courant-Snyder $\beta$-function.
Both ion species are assumed to have Gaussian momentum distributions with similar rms momentum spreads of
$\delta_{\mathrm{rms}} \approx \num{0.5e-3}$~\cite{holzfeind2024implementation}.
For this illustration, both ion species are further modeled with Gaussian betatron distributions with a normalized rms emittance of
$\epsilon_{x,\mathrm{n,rms}} = \SI{0.9}{\micro\meter}$~\cite{pivi_status_2019}.
While in practice, the horizontal emittances of the helium and carbon distributions differ significantly when generating the mixed beam using a double multiturn injection scheme~\cite{kausel_prab},
this distinction is not relevant for the present analysis and is therefore omitted here for simplicity.
The bottom row shows the respectively modeled horizontal charge densities at the location of the radia loop pick-up. 

\subsection{Radial loop regulation of mixed-species beams}
\begin{figure*}[tb]
    \centering
    \includegraphics[    
    width=\linewidth]{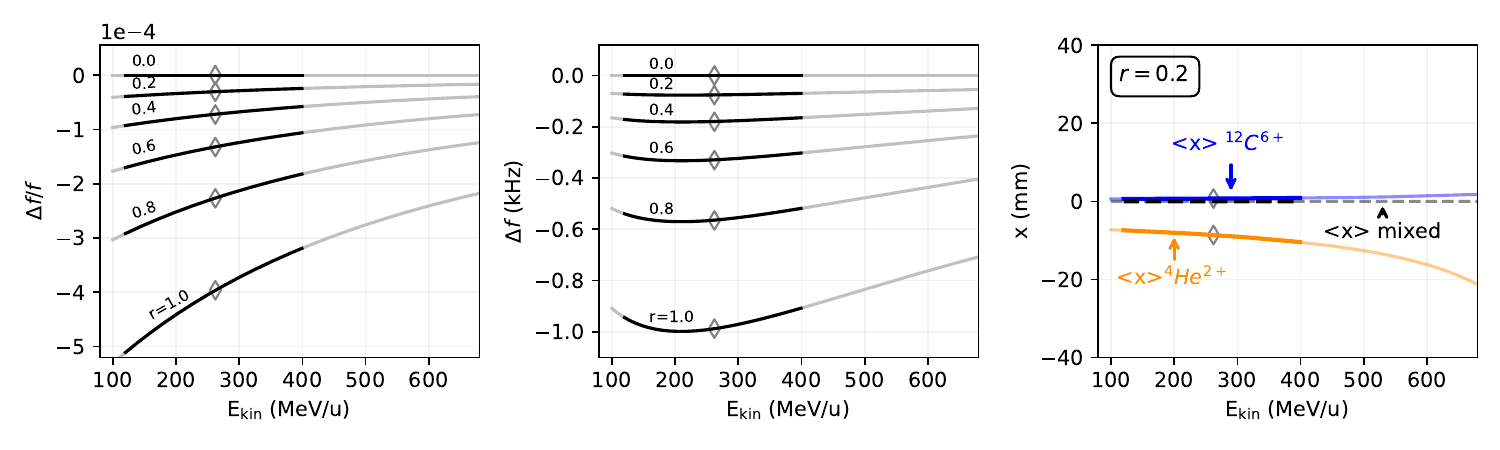}
    \vspace{-0.8cm}
    \caption{\raggedright Frequency and radial position offsets for mixed helium-carbon beams with different ion fractions $r$, estimated analytically (lines) and in simulations (grey diamond markers). The clinically relevant energy range is indicated by enhanced opacity. It should be noted that the present formalism is not expected to hold in the vicinity of transition energy (here $\gamma_\mathrm{tr}\approx 2$). Left: relative frequency offsets with respect to a mono-isotopic \csix{} beam. Center: Absolute frequency offsets in a PIMMS synchrotron. Right: Horizontal centroid of the mixed, carbon, and helium charge distributions at the radial loop pick-up with ${D_x = \SI{-8.3}{\metre}}$. In this example the radial loop regulates to ${x_\mathrm{RL}=\SI{0}{\milli\metre}}$. }
    \label{fig:expected_dff}
\end{figure*}

The radial loop acts on the overall charge distribution (black-dashed line in the bottom row of \cref{fig:schematic_deltaR}) and applies a frequency correction $\Delta f(r) = f_\mathrm{mix}(r) - f_\mathrm{ref}$  to move $\left\langle x\right\rangle_\mathrm{q,mix}$ to the specified radial loop setpoint. $f_\mathrm{mix}$ denotes the RF frequency of the mixed beam after regulation  to the prescribed offset $x_\mathrm{RL}$. $f_\mathrm{ref}$ is the corresponding reference frequency of a mono-isotopic beam of the reference ion species (only species~I, i.e.\ $r = 0$), which is acquired at the same radial loop setpoint. In the remaining section, the subscript ``ref'' is omitted when describing the  reference beam, i.e.\ ${f = f_\mathrm{ref}}$, ${\gamma = \gamma_\mathrm{ref}}$, ${\beta = \beta_\mathrm{ref}}$, ${p = p_\mathrm{ref}}$ and ${R = R_\mathrm{ref}}$.

After regulation, the displacement of the reference species~I with respect to the mono-species reference position follows from \cref{eq:avg_x_q_mix} as
\begin{equation}
    x_\mathrm{I}\left(r\right) - x_\mathrm{RL} = -  A_\mathrm{r} \, D_x \, \delta_\mathrm{eff,sp}~.
\end{equation}
This offset can be related to the induced momentum shift of species~I, ${\Delta p_\mathrm{I} = p_\mathrm{I} - p}$, by
\begin{equation}
    D_x \, \dfrac{\Delta p_\mathrm{I}}{p} = -  A_\mathrm{r} \, D_x \, \delta_\mathrm{eff,sp}~.
\end{equation}
$D_x$ cancels out, assuming it is approximately constant over the relevant range. Finally, the obtained $\Delta p_\mathrm{I}/p$ can be related to the induced frequency deviation,
\begin{equation}\label{eq:dff_multiply_eta_deff}
    \dfrac{\Delta f}{f} =\eta \, \dfrac{\Delta p}{p} =- \eta\, A_\mathrm{r} \,  \, \delta_\mathrm{eff,sp}~,
\end{equation}
with the slippage factor $\eta = \frac{1}{\gamma^2}  - \frac{1}{ \gamma_\mathrm{tr}^2}$. Rearranging with \cref{eq:hat_delta,eq:delta_eff_lin} yields the relative frequency offset expected for a specific ion fraction $r$,
\begin{equation}\label{eq:dff_for_r_new}
    \dfrac{\Delta f}{f} = - \dfrac{1}{\gamma^2}  \, A_\mathrm{r} \left(\dfrac{1}{\chi_\mathrm{sp}}-1\right)~.
\end{equation}
Equivalently, the ion fraction $r$ can be obtained from a measured $\Delta f/f$ by rearranging \cref{eq:dff_for_r_new}, yielding
\begin{equation}\label{eq:r_for_dff_new}
        r = \dfrac{
            q_\mathrm{I} \, \gamma^2 \, \dfrac{\Delta f}{f}
            }{\gamma^2 \, \dfrac{\Delta f}{f} \, \left(q_\mathrm{I}-q_\mathrm{II}\right) - q_\mathrm{II}\, \left(\dfrac{1}{\chi_\mathrm{sp}}-1\right)}.\\
\end{equation}

Notably, \cref{eq:dff_for_r_new,eq:r_for_dff_new} remain independent of the machine layout, as the dependence on the momentum compaction factor cancels out in \cref{eq:dff_multiply_eta_deff}.

Alternatively, the situation can also be conceptualized by means of a virtual particle representing the charge-weighted center of gravity of the mixed-beam distribution. 
This virtual particle is described by the effective charge-to-mass ratio factor $\chi_\mathrm{v}$, which is obtained as charge-weighted average of $\frac{1}{\chi}-1$,
\begin{equation}
    \begin{aligned}
    \dfrac{1}{\chi_\mathrm{v}}-1 &=  \dfrac{\left(
            \frac{1}{\chi_\mathrm{sp}}-1\right)\, q_\mathrm{II}\, r 
        }{
            q_\mathrm{II}\, r + q_\mathrm{I}\, (1-r)
        }
\end{aligned}
\end{equation}
consistent with the fact that the change in dispersive offset is linear in $\frac{1}{\chi} - 1$, see \cref{eq:dx_he_c}. Using $\frac{1}{\chi} -1 \approx 1 -\chi$, we obtain
\begin{equation}\label{eq:averagechi}
    \begin{aligned}
    \chi_\mathrm{v} &=  \dfrac{
            \chi_\mathrm{sp}\, q_\mathrm{II}\, r + q_\mathrm{I}\, (1-r)
        }{
            q_\mathrm{II}\, r + q_\mathrm{I}\, (1-r)
        }~.
    \end{aligned}
\end{equation}

Also this approach requires that $D_x$ is approximately constant over the relevant range. 

The radial loop regulates the dispersive offset of this virtual particle to the same position as the mono-isotopic reference beam. This means, that the effective rigidity offset of the virtual particle with respect to the reference beam vanishes,
\begin{equation}
    \delta_\mathrm{eff,v} = \dfrac{1}{\chi_\mathrm{v}} -1 + \delta_\mathrm{v}= 0,
\end{equation}
which requires the virtual particle to feature a relative offset in $\beta\gamma$ of
\begin{equation}\label{eq:delta_v_chi_rel}
    \delta_\mathrm{v} =1 - \dfrac{1}{\chi _\mathrm{v}} \approx \chi _\mathrm{v}-1.
\end{equation}

Since the radial loop aligns the center of gravity of the mixed beam with the reference orbit, the effective path length of the virtual particle equals that of the reference beam, i.e. $R_\mathrm{v} = R$ in \cref{eq:dff}. Using \cref{eq:beta_expansion}, the relative frequency offset becomes
\begin{equation}\label{eq:dff_v}
    \begin{aligned}
    \dfrac{\Delta f}{f} &  = \dfrac{\beta_\mathrm{v}}{\beta} \dfrac{R}{R_\mathrm{v}} -1 = \dfrac{1}{\gamma^2} \,\delta_\mathrm{v}+ \mathcal{O}\left(\delta_\mathrm{v}^2\right)~.
    \end{aligned}
\end{equation}

The measured relative frequency shift $\Delta f/f$ can therefore be used to estimate the relative offset in $p/m = \beta \gamma c$ of the virtual particle,
\begin{equation}\label{eq:deltav_from_dff}
        \delta_\mathrm{v} \approx \gamma^2 \dfrac{\Delta f}{f}.
\end{equation}
With \cref{eq:averagechi,eq:delta_v_chi_rel} one obtains again \cref{eq:dff_for_r_new,eq:r_for_dff_new}, as derived above. The independence from the machine layout stems in this approach from postulating $R_\mathrm{v} = R$, which eliminates any effective path length difference for the virtual particle. 

To provide an example, \cref{fig:expected_dff}    illustrates the resulting relations for a mixed \hetwo{} and \csix{} beam
in a PIMMS synchrotron lattice, with a pure \csix{} beam as the reference species. 
The example explores energies between \SI{120}{\mega\electronvolt\per\u} and \SI{680}{\mega\electronvolt\per\u} and ion fractions between ${r=0}$ (pure \csix{}) and ${r=1}$  (pure \hetwo{}). While this example assumes on-momentum operation for simplicity, the relations remain valid for off-momentum operation. 

Notably, it is evident in \cref{fig:expected_dff} (left and center panels) that the method is not applicable for ${\gamma \gg 1}$ as the induced frequency offset vanishes in accordance with the $1/\gamma^2$ dependence in \cref{eq:dff_for_r_new}. 
Additionally, we emphasize that the present formalism is not expected to apply in the vicinity of transition, which lies beyond the scope of this work.

To validate \cref{eq:dff_for_r_new,eq:r_for_dff_new}, simulations are performed with \textsc{Xsuite}~\cite{iadarola_xsuite_2024}, which enables tracking of particles with non-nominal mass-to-charge ratios. Details are outlined in \cref{sec:app:benchmark}. In these simulations, $\Delta f$ is determined numerically by enforcing the charge-weighted average radial displacement ${\langle x \rangle_\mathrm{q,mix}}$ to be ${x_\mathrm{RL}}$. As substitute for a radial feedback loop, the required frequency offset $\Delta f$ is determined using the Nelder–Mead optimization algorithm~\cite{NelderMead}, implemented in \textsc{SciPy}~\cite{2020SciPy-NMeth}.  The resulting frequency offsets and the corresponding radial displacements are shown as gray diamond markers in \cref{fig:expected_dff}, in excellent agreement with the expected values.


\section{Experimental demonstration}\label{sec:experiment}

The proposed method for estimating the ion fraction of dual-species beams was experimentally tested in 2025 at the MedAustron center for ion beam therapy and research.

\subsection{Setup and measurement campaign}

\begin{figure}
    \centering
    \includegraphics[    
    width=\linewidth]{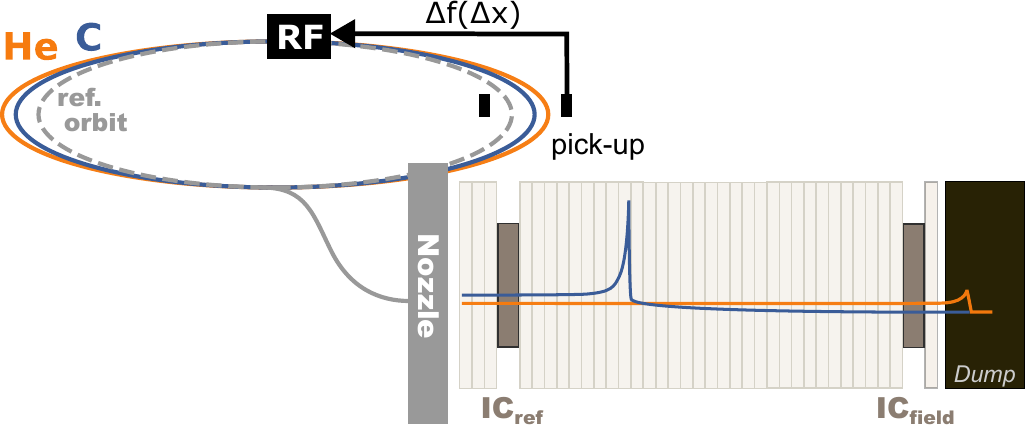}
    \caption{Schematic of the two complementary methods to determine the ion composition of the dual-species beam: the noninvasive RF-based measurement in the synchrotron and the post-extraction measurement with an ionization-chamber (IC) telescope.}
    \label{fig:schematic_IC}
\end{figure}

For this purpose, the helium fraction of mixed \hetwo{} and \csix{} beams was determined both at flat top in the synchrotron (\cref{ch:campaign:rf}) and independently after extraction in the research irradiation room (\cref{ch:campaign:ionizationchamber}). The setup is schematically illustrated in \cref{fig:schematic_IC}.
Measurements were performed at two extraction energies, \SI{262.3}{\mega\electronvolt\per\u} and \SI{402.8}{\mega\electronvolt\per\u}. 
For  \SI{262.3}{\mega\electronvolt\per\u}, measurements were carried out with two distinct radial loop setpoints, ${x_\mathrm{RL}=\SI{27.5}{\milli\metre}}$ and ${\SI{29.9}{\milli\metre}}$, to further verify the robustness of the measurement procedure.
The radial loop offset ${x_\mathrm{RL}=\SI{27.5}{\milli\metre}}$ represents the default setting at flat top in the employed extraction setup (\cref{ch:campaign:mixedbeamdelivery}). 
The second setpoint, $x_\mathrm{RL} = \SI{29.9}{\milli\metre}$, was selected to be sufficiently close to the initial setpoint to avoid significantly affecting the dynamics during slow extraction. The specific value is not important for the conclusions presented here. It was chosen solely to reproduce the relative frequency offset used in previous tests at a different energy (data not shown).
For each energy and radial loop configuration, 
the ion fraction was varied by means of adapting the sequential injection settings (\cref{ch:campaign:mixedbeamdelivery}). For each setting, five measurements were taken. 

\subsubsection{Mixed-beam delivery}\label{ch:campaign:mixedbeamdelivery}

The mixed beams were produced using the double multiturn injection scheme described in \cref{sec:mixedbeam_gen_medaustron}. Different ion fractions were obtained by varying the amplitude of the second injection bump, which is used for the injection of \csix\ ions. As a full calibration of the bump amplitude was not available at the time of writing, the bump strength is specified in terms of the injection bump current. By adjusting the injection kicker current between ${I_\mathrm{MKI}=\num{350}}$ and \SI{0}{\ampere} ion fractions ranging from ${r=0}$ (pure \csix{}) to ${r=1}$ (pure \hetwo{}) were achieved.

Slow extraction to the experimental room was performed by means of phase displacement acceleration~\cite{dorda:ipac13-wepea001,ARRUTIASOTA2022167007,renner_investigating_2024}, in which the beam is pushed into the third-order resonance by sweeping an empty RF bucket through a coasting beam stack. This extraction mechanism was used because the clinically employed betatron core is currently not compatible with mixed-species extraction, and the experimentally available RF knockout system was not available for the extraction at the time of the measurements~\cite{renner:ipac2026-thp4091}.

For all measurements, the extraction parameters were chosen to extract the beam over ${\approx\SI{7}{\second}}$ while ensuring complete emptying of the synchrotron and avoiding saturation of the ionization chambers in the research irradiation room. For both beam energies, the empty RF bucket was empirically initialized at a frequency offset that did not induce beam losses upon application of RF voltage. A combination of RF voltage, i.e.\ bucket height, and frequency sweep speed was then empirically selected to empty the entire synchrotron over several seconds.

The extraction efficiency was estimated for pure \csix{} beams by comparing intensity measurements in the synchrotron with readings from the ionization chambers in the nozzle of the MedAustron research irradiation room. This yielded extraction efficiencies of $96.1\pm \SI{0.7}{\percent}$ for \SI{262.3}{\mega\electronvolt\per\u} with a \SI{27.5}{\milli\metre} radial loop offset, $95.6\pm\SI{0.6}{\percent}$ for \SI{262.3}{\mega\electronvolt\per\u} with a \SI{29.9}{\milli\metre} offset, and $96.0\pm\SI{0.2}{\percent}$ for \SI{402.8}{\mega\electronvolt\per\u} with a \SI{25}{\milli\metre} offset.
However, operation in mixed-beam mode may lead to species-dependent changes in extraction efficiency relative to the reference measurements obtained with a pure carbon beam, which should therefore be taken into account when interpreting the results. 

\subsubsection{Measuring the ion composition in the synchrotron}\label{ch:campaign:rf}

\begin{figure}
    \includegraphics[width=\linewidth]{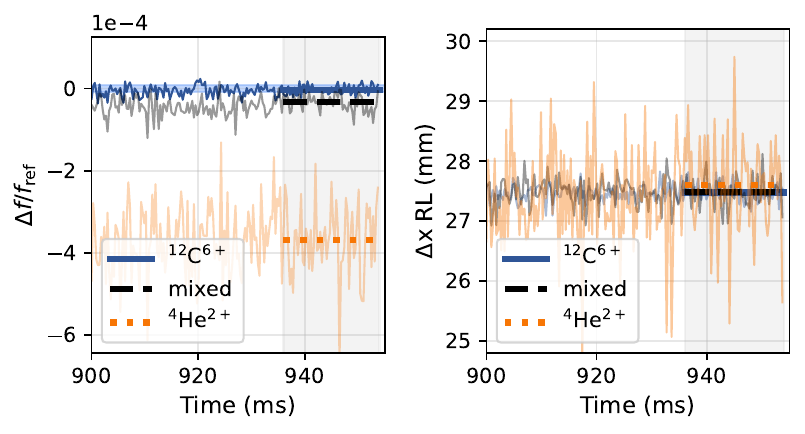}
    \caption{
    LLRF acquisitions at flat top for \SI{262.3}{\mega\electronvolt\per\u} \csix{}, \hetwo{}  and mixed beams ($r\approx0.2$), with the radial loop regulating to ${x_\mathrm{RL} = \SI{27.5}{\milli\metre}}$. The gray-shaded time window is used to determine the average frequencies, shown by the solid lines.
    Left: Relative RF frequency offset with respect to a pure \csix{} beam.
    The blue-{\allowbreak}shaded band marks the reference
    frequency $f_\mathrm{ref} \pm 3\sigma_{f_\mathrm{ref}}$.
    Right: Position measured at the radial loop pick-up.
    }
    \label{fig:freq}
\end{figure}

\Cref{fig:freq} shows representative low-level RF (LLRF)
acquisitions used for the RF-based ion fraction quantification. 
As expected, the mixed \hetwo{}/\csix{} beam (black) exhibits a frequency offset that lies between those of the two pure species (orange and blue).
The  frequency values required for the mixing-ratio quantification were  obtained by averaging the acquired RF frequency signal over a fixed time interval during the flat-top period prior to extraction (grey-shaded region in \cref{fig:freq}). In the presented measurements, these intervals were 18\,ms for \SI{262.3}{\mega\electronvolt\per\u} beams and 6\,ms for \SI{402.8}{\mega\electronvolt\per\u} beams. These time windows were selected as the longest intervals at the end of the LLRF acquisition during which the frequency averages remained unaffected by residual magnetic field settling. The intervals are relatively short for the data presented because the main dipole magnets were operated in current-regulated mode, and the magnetic field had not yet fully stabilized by the end of the LLRF acquisition.

For each configured energy and radial loop setpoint, the reference frequency $f_\mathrm{ref}$ ($f$ in \cref{eq:dff_for_r_new})  and corresponding standard error $\sigma_{f_\mathrm{ref}}$  were determined at the beginning of the measurement campaign by averaging 5~independent acquisitions with a pure \csix{} beam. The obtained rms frequency spread is $\sigma_{f_\mathrm{ref}} < \qty{8}{\hertz}$ for all extraction energies and radial loop offsets, which is therefore adopted as common uncertainty for all cases. This spread is mainly attributed to noise in
 the RF frequency read-out and to magnetic field fluctuations of the
main dipole magnets. The blue-shaded band in the left subplot of \cref{fig:freq} represents the resulting \csix{} reference frequency interval of $\pm 3\sigma_{f_\mathrm{ref}}$. 

For each subsequent mixed-beam measurement, the revolution
frequency $f_\mathrm{mix}$ is measured, the corresponding frequency shift
$\Delta f = f_\mathrm{mix} - f_\mathrm{ref}$ calculated, and the ion
fraction $r$ obtained via \cref{eq:r_for_dff_new}.

\Cref{fig:freq} further shows that the frequency fluctuations are noticeably larger for the pure
\hetwo{} beam than for the mixed and pure \csix{} beams.
This is attributed to the lower helium beam current
and to the gain setting of the pick-up amplifier, which is optimized
for \csix{} operation. This results in less
stable regulation for beams with ${r \rightarrow 1}$. Nevertheless, the measurement quality is sufficient for the presented
proof-of-principle demonstration. 
Importantly, the pick-ups installed in the MedAustron synchrotron are of the shoe-box type, which provide a linear relationship between beam displacement and induced signal over a wide displacement range~\cite{Strehl:2006sho}, as required for the presented model.

Representative parameter sets and the corresponding rms uncertainties used for error propagation are listed in \cref{tab:errors}. The uncertainty of the extraction energy is conservatively assumed to be 3\,MeV/u to account for non-nominal synchrotron operation in the mixed-beam setup. However, its effect is negligible, as the uncertainty in $\Delta f$ is the dominant contribution to the uncertainty in $r$.
\begin{table}
    \centering
    \caption{Parameters for the helium fraction quantification using the  RF-based method in the synchrotron and the ionization chamber telescope in the irradiation room.}
    \label{tab:errors}
    \renewcommand{\arraystretch}{1.15}
    \begin{tabular}{l c c c}
        \hline
        \hline
        Parameter & Value & Standard error & Unit \\ 
        \hline \\ [-3pt]
        \multicolumn{4}{l}{\textit{263.2\,MeV/u \quad 27.5\,mm radial offset}} \\ [3pt]
        $f_\mathrm{ref}$& \num{2.405142e6} & \num{8} & Hz \\
        $\Delta f$    & $\Delta f(r)\in[0,-1000]$ & \num{8} & Hz \\
        $E_\mathrm{kin}$ & 263.2 & 3 &  MeV/u \\
        $a_\mathrm{He}$ & 1.54 & \num{4e-3} &  \\
        $a_\mathrm{C}$  & 0.01 & \num{2e-5} &  \\
        $\Delta E_\mathrm{He}$ & \num{5.69e-3} & \SI{5}{\percent} & MeV  \\
        $\Delta E_\mathrm{C}$ & \num{5.45e-2} & \SI{5}{\percent} & MeV  \\
        \\ [-3pt]
        \multicolumn{4}{l}{\textit{402.8\,MeV/u \quad  25\,mm radial offset}}\\ [3pt]

        $f_\mathrm{ref}$& \num{2.757010e6} & \num{8} & Hz \\
        $\Delta f$    & $\Delta f(r)\in[0,-900]$ & \num{8} & Hz \\
        $E_\mathrm{kin}$ & 402.8 & 3 &  MeV/u \\
        $a_\mathrm{He}$ & 0.77 & \num{2e-3} &  \\
        $a_\mathrm{C}$  & 0.15 & \num{5e-4} &  \\
        $\Delta E_\mathrm{He}$ & \num{4.53e-3} & \SI{5}{\percent} & MeV  \\
        $\Delta E_\mathrm{C}$ & \num{4.07e-2} & \SI{5}{\percent} & MeV  \\
        \hline
        \hline
    \end{tabular}
\end{table}

\subsubsection{Measuring the ion composition of the extracted beam}\label{ch:campaign:ionizationchamber}

\paragraph{Methodology:} In the research irradiation room, a two-stage ionization chamber (IC) telescope was used to quantify the mixing ratio of the extracted beam. As illustrated in \cref{fig:schematic_IC}, the two ICs are positioned at different penetration depths, with the \textit{reference chamber} $\mathrm{IC}_\mathrm{ref}$ being located in the entrance plateau of the carbon Bragg peak, while the \textit{field chamber} IC$_\mathrm{field}$ is located in between the \csix and the \hetwo Bragg peaks. The ion composition can be reconstructed from the ratio of the currents measured simultaneously on both ICs, $I_\mathrm{ref}$ and $I_\mathrm{field}$~\cite{kausel:ipac2025-tupb099}.

The method relies on calibration factors that relate the average signal induced by a primary \hetwo{} or \csix{} ion, or by their fragments, in IC\textsubscript{field} to the corresponding signal in IC\textsubscript{ref}. These factors are determined in dedicated calibration measurements performed at the beginning of the experiment by irradiating the IC telescope with pure helium and pure carbon beams and computing
\begin{equation}\label{eq:afactors}
    \begin{aligned}
    a_j &= \dfrac{I_{\mathrm{field},j}}{I_{\mathrm{ref},j}} \qquad {j=\text{He and C}}~.\\
\end{aligned}
\end{equation}
To first order, these calibration factors are independent of the incident beam intensity.

During measurements with the mixed-species beam, the total current in the two ICs is
\begin{equation}
    \begin{aligned}
        I_{\mathrm{ref}}(t) &= I_{\mathrm{ref,He}}(t) + I_{\mathrm{ref,C}}(t)\\
        I_{\mathrm{field}}(t) &= I_{\mathrm{field,He}}(t) + I_{\mathrm{field,C}}(t)~,
    \end{aligned}
\end{equation}
where $t$ describes the time during the $\approx \qty{10}{\second}$ spill. Using the calibration factors defined in \cref{eq:afactors}, the current in the field IC can be written as
\begin{equation}
        I_{\mathrm{field}}(t) = a_{\mathrm{He}} \, I_{\mathrm{ref,He}}(t)
                    + a_{\mathrm{C}} \, I_{\mathrm{ref,C}}(t).
\end{equation}
Solving this system of equations yields 
\begin{equation}\label{eq:lin_sys_eq}
    \begin{aligned}
        I_\mathrm{ref,He}(t)  &= \dfrac{I_\mathrm{field}(t)  - a_\mathrm{C}\,I_\mathrm{ref}(t)}{a_\mathrm{He}-a_\mathrm{C}}\\
        I_\mathrm{ref,C}(t)  &= \dfrac{a_\mathrm{He} \, I_\mathrm{ref}(t)  - I_\mathrm{field}(t)}{a_\mathrm{He}-a_\mathrm{C}}.\\
    \end{aligned}
\end{equation}
The integrated reference IC current for each species~$j$ over the spill duration $T$ is
\begin{equation}
    I_{\mathrm{ref},j} = \sum_{i=0}^{T/\Delta t}{I_{\mathrm{ref},j}(t_i)\Delta t}~,
\end{equation}
where $\Delta t$ is the sampling time of the electrometer. 
The fraction of extracted ions per species is subsequently obtained from the ratio of these integrated currents and the ratio $\Delta E_\mathrm{He}/\Delta E_\mathrm{C}$ of the energy deposited in the detector volume per primary \hetwo{} and \csix{} ion for the given experimental configuration,
\begin{equation}\label{eq:integrat_NheNC}
    \frac{N_\mathrm{He}}{N_\mathrm{C}} =
    \frac{I_{\mathrm{ref,He}}}{I_{\mathrm{ref,C}}}
    \frac{\Delta E_\mathrm{C}}{\Delta E_\mathrm{He}}.
\end{equation}
The ion fraction $r$ then follows as
\begin{equation}
    r = \dfrac{\frac{N_\mathrm{He}}{N_\mathrm{C}} }{\frac{N_\mathrm{He}}{N_\mathrm{C}} + 1}.
\end{equation}
$\Delta E_\mathrm{He}/\Delta E_\mathrm{C}$ is obtained from \textsc{Opengate}~\cite{sarrut_opengate_2022} simulations, with details described below. 

\vspace{5mm}
\paragraph{Details on the experimental setup:}
The presented experiment employed PTW 34070 ICs, separated in range by multiple PTW RW3 absorber plates (PTW Freiburg GmbH, Freiburg, Germany). The reference chamber was positioned after 3~plates, with another 33~plates installed in between reference and field IC. Each plate has a physical thickness of 1\,cm. Their water equivalent thickness has been characterized with protons and is  available internally at the MedAustron facility. Both ICs fully covered the entire beam, which was verified using Gafchromic\texttrademark\ EBT3 radiochromic films (Ashland Inc., Wayne, NJ, USA) upstream and downstream of the IC telescope. The chosen configuration assured that the carbon Bragg peak position (around 14\,cm for \SI{262.3}{\mega\electronvolt\per\u} and 28\,cm for \SI{402.8}{\mega\electronvolt\per\u}) was in between the two ICs for both energies while the \hetwo{} Bragg peak (approximately 41\,cm for \SI{262.3}{\mega\electronvolt\per\u} and 83\,cm for \SI{402.8}{\mega\electronvolt\per\u}) was located after the field IC. 
The ICs were read out with a 50\,ms integration time using a PTW TANDEM dual channel electrometer.

The setup-specific parameters and the corresponding rms uncertainties used for error propagation are summarized in \cref{tab:errors}. The calibration factors $a_\mathrm{He}$ and $a_\mathrm{C}$ were obtained by averaging five acquisitions taken at the beginning of the measurement campaign. To obtain $\Delta E_\mathrm{He}$ and $\Delta E_\mathrm{C}$, a simplified model of the setup was simulated in \textsc{Opengate} (version 10.0.3 using \textsc{Geant4}~\cite{agostinelli_geant4simulation_2003} version 11.2.3). The PTW RW3 slabs and the ICs were implemented explicitly according to their material composition and internally available data for their water equivalent thickness. The active region of the ICs was modeled as \qty{2}{\milli\metre} air-filled volumes with energy deposition scored over the full sensitive detector thickness and cross section. 
The simulations employed the \textsc{Geant4} \textsc{Shielding\_EMZ} physics list~\cite{geant4_physlist_2025}, which provides stopping-power data based on the ICRU Report 90~\cite{ICRU90} for protons, alpha particles, and heavy ions in water, graphite, and air. Given the high energies of the helium and carbon ions and the low fragmentation upstream of the reference IC, the ICRU90 stopping-power data are expected to dominate the energy-loss modeling in the detector volumes. This study assumes an rms error of $\SI{5}{\percent}$ for the estimate on the deposited energy $\Delta E$. On the one hand, this value contains systematic uncertainties related to stopping power data, which are assumed to be $\pm\SI{2.5}{\percent}$ in accordance with recent comparisons of stopping powers obtained with different \textsc{Geant4} physics lists~\cite{Lee2025StoppingPower}. Further, the error budget accounts for statistical simulation uncertainties (${<\SI{1}{\percent}}$ ) and systematic errors in the simulation setup. A more precise quantification of these uncertainties is subject to future studies and goes beyond the presented proof of principle. 

\subsubsection{Data analysis}\label{ch:campaign:dataanalysis}

To quantify potential systematic differences between the two independent ion fraction estimates, a simple phenomenological model is introduced.
For each method the inferred helium fraction is written as
\begin{equation}
    r_{i} = \frac{\lambda_{i} N_{\mathrm{He}}}{\lambda_{i} N_{\mathrm{He}} + N_{\mathrm{C}}},\qquad{} i=\mathrm{ic, rf}
\end{equation}
where $N_{\mathrm{He}}$ and $N_{\mathrm{C}}$ denote unknown ground truth, i.e. the intensities prior to both measurements. $\lambda_i$ accounts for  
method-specific response differences, such as species-dependent quantification or
extraction efficiencies.

Eliminating $N_{\mathrm{He}}$ and $N_{\mathrm{C}}$ yields a direct relation between the ion fraction 
\begin{equation}
    \kappa = \frac{\lambda_\mathrm{rf}}{\lambda_\mathrm{ic}},
\end{equation}
the expected relation reads
\begin{equation}\label{eq:modelkappa}
    \widehat{r}_\mathrm{rf}(r_\mathrm{ic}) =
    \frac{\kappa\, r_\mathrm{ic}}{1 - r_\mathrm{ic} + \kappa\, r_\mathrm{ic}}.
\end{equation}

In the presented measurements, the parameter $\kappa$ was determined by ordinary least-squares regression. Pure-species points (${r=0}$, ${r=1}$) are, by design, detected as such by the IC setup, independent of detection and extraction efficiencies. These data points were excluded from the fit. The fitted parameter $\kappa$ represents an effective relative response between the two methods, incorporating differences in detection and extraction efficiencies for the presented measurement setup. The adequacy of this model was assessed using the residuals $r_\mathrm{rf} - \widehat{r}_\mathrm{rf}(r_\mathrm{ic})$.

\subsection{Experimental results}
\begin{figure*}
    \includegraphics[width=\textwidth]{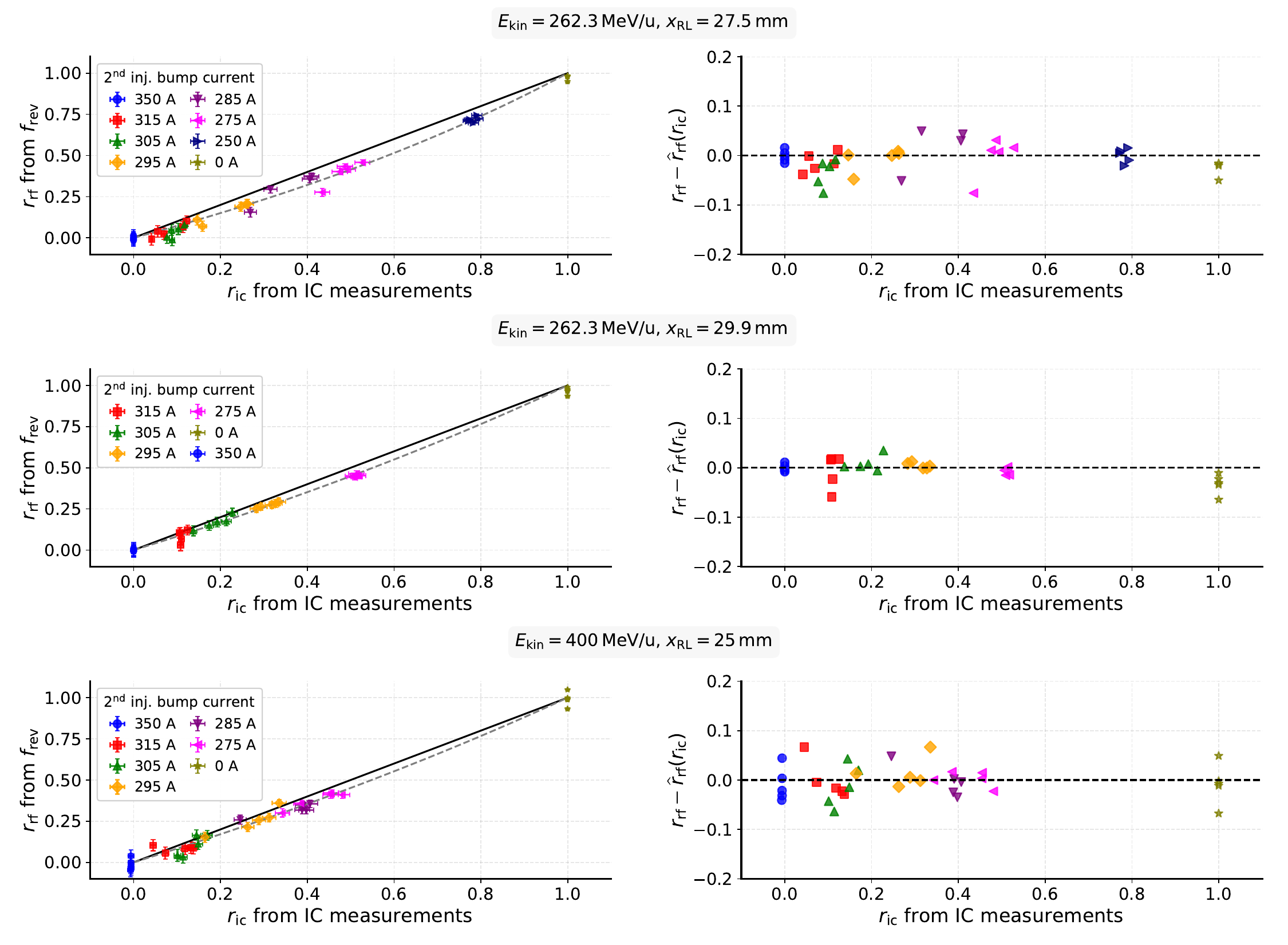}
    \caption{Left: Comparison of the ion fraction obtained with the ionization chamber telescope in the irradiation room (x-axis) and the noninvasive RF-based quantification in the synchrotron (y-axis) for different beam energies and radial loop offsets. The black solid line indicates the identity ${r_\mathrm{ic}=r_\mathrm{rf}}$, the gray dashed line shows the linear fit based on the efficiency model. Right: Residuals of the fit.}
    \label{fig:mixing_ratios_comparison}
\end{figure*}

The ion fractions $r$ estimated using the two different methods are compared in
\cref{fig:mixing_ratios_comparison}.
The left panels show a direct comparison between the ion fraction
of the extracted beam measured in the research irradiation room (x-axis, $r_\mathrm{ic}$),
and the noninvasive composition estimate obtained in the synchrotron
(y-axis, $r_\mathrm{rf}$).
The colored markers correspond to different injection settings used to tailor the ion fraction. 

For a fixed injection setting, i.e.\ one color in \cref{fig:mixing_ratios_comparison}, the measured ion fractions exhibit significant fluctuations in selected cases.
The fact that these variations are consistently reflected in both methods indicates that
these are real shot-to-shot fluctuations of the beam composition rather than measurement artifacts. 

Importantly, the pure-species limit ${r=1}$, which is intrinsically identified as such by the IC-based measurement, is also recovered as ${r \approx 1}$ by the RF-based method. This agreement provides a strong consistency check of the underlying model, as it follows directly from the calculated frequency offset rather than from any interpolation.

At intermediate ion fractions, deviations between the two methods are observed.
The fit of the model in \cref{eq:modelkappa} yields a value of
${\kappa \approx \qtyrange{0.75}{0.83}{}}$ for the different measurement
sets.
This indicates a difference in the relative response to the initially present \hetwo{} and
\csix{} between the two methods, which may arise from species-dependent detection
efficiencies and/or differences in extraction efficiency. The deviations are largest at intermediate ion fractions ($r\approx0.5$), where $\kappa\approx0.8$ corresponds to the in-synchrotron measurement yielding a helium fraction that is approximately up to $\SI{6}{\percent}$ lower than that inferred from the ICs in the experimental room. 
However, as the present results serve as a proof of principle, the regression is used only to assess systematic trends. A detailed quantitative and physical discussion of the fit parameters is deferred to future studies.

The residuals of the fitted model are shown in the right panels. For all investigated configurations, the residuals are centered around zero and remain within approximately $\pm\SI{7}{\percent}$, with no significant systematic trends observed over the full range. This supports  the validity of the model parameterized by $\kappa$, \cref{eq:modelkappa}.


\section{Discussion and outlook}\label{sec:discussionoutlook}

\subsection{Method applicability and limitations}

The presented results demonstrate the feasibility of noninvasively estimating the ion composition of dual-species beams directly in the synchrotron, with good agreement between the RF-based method and the IC telescope. While the current implementation is not optimized for high-precision measurements, it already provides significant practical value and is regularly used for mixed-beam commissioning. Further, for experiments where the installation of an IC telescope was incompatible with the experimental setup, the presented noninvasive approach enabled cycle-resolved composition monitoring without interfering with beam delivery.

Looking ahead to potential clinical implementation, the results indicate that this measurement principle could possibly be used to monitor the ion composition during treatment and generate interlocks in the case of unacceptable mixing ratios, similar to an energy verification system.  Further studies are recommended to assess the feasibility of this approach and investigate concepts for directly integrating the ion fraction determination into the LLRF system. For these studies, input from medical physics research groups is needed to define the required accuracy and precision, which are expected to be more stringent than those currently achieved.

The achievable accuracy and precision of the RF-based method, as well as their evaluation, are currently limited by several factors. Regarding accuracy, an assessment remains challenging due to the absence of a ground truth. So far, the helium fraction estimated via the RF-based method could only be compared with the IC measurement in the irradiation room. 
However, this comparison may be affected by differences in extraction efficiency between the two ion species, as well as  uncertainties in the IC-based reconstruction, both of which require further investigation.

Regarding precision, the key constraints in this study were fluctuations in the acquired frequencies, with $\frac{\Delta f}{f}$ at the $\mathcal{O}(10^{-6})$ level. These fluctuations are attributed mainly to noise in the LLRF acquisition system and cycle-to-cycle magnetic field fluctuations~\cite{margreiter:ipac2024-thpg16}.

The LLRF acquisition noise is expected to be reducible by extending the sampling duration. For example, averaging over several tens of milliseconds instead of just a few milliseconds is expected to reduce the rms error due to LLRF noise by a factor of 2--3, following the dependence $\sigma_\mathrm{rms} \propto 1/\sqrt{n_\mathrm{s}}$ on the sample size $n_\mathrm{s}$. If simple averaging is applied to retrieve the frequency, longer sampling times are only feasible if the magnetic field remains stable during the measurement period. In the presently employed current-regulated magnetic field control, this requires extending the flat-top duration by $\approx \SI{2}{\second}$  to ensure that the field is fully settled prior to the measurement~\cite{Margreiter:2025pkh}. This is a straightforward mitigation measure for the research setup and is therefore foreseen for future mixed-beam campaigns. However, it is not suitable for potential clinical application, where significantly prolonged cycle times are unacceptable. Alternatively, longer sampling times without an extended flat top could be enabled by fitting a smooth model function to the time evolution of the drifting frequency or by operating the main dipoles in $B$-field-regulated mode~\cite{Margreiter:2025pkh}.

The second factor that limits measurement precision  is cycle-to-cycle $B$-field fluctuations, which have been assessed to be $\frac{\Delta B}{B} \lesssim 10^{-5}$ for both current- and $B$-field-regulated modes~\cite{margreiter:ipac2024-thpg16,Margreiter:2025pkh}. Using the common differential relations between $B$-field, frequency and mean orbit radius~\cite{Bovet:104153}, this corresponds to a frequency uncertainty $\frac{\Delta f}{f}$ at the $\mathcal{O}(10^{-6})$ level. While these fluctuations therefore clearly limit the precision of the proposed method, they are below the tolerances required for nominal clinical operation, and significant further reduction is not expected in current facilities.

Generally, when aiming to increase precision, the regime with ${r\approx \qtyrange{0.1}{0.4}{}}$ should be prioritized, as this is the range of helium fractions most commonly requested by experiments. The reduced signal-to-noise ratio in the LLRF acquisitions for high helium fractions ($r\rightarrow 1$, orange in \cref{fig:freq}), which arises because the gain settings are optimized for carbon, is therefore not considered a significant issue within the relevant operational range.

\subsection{Operational implications of radial loop regulation for mixed-species beams}

The frequency correction applied by the radial loop for a given ion composition, \cref{eq:dff_for_r_new}, induces a composition-dependent shift in $\beta\gamma$ and, consequently, also in the flat-top energy of the reference ion species used for tumor treatment. When the beam is extracted using  amplitude-driven slow extraction (RF knock-out), this energy offset is also maintained during  extraction and results in a corresponding range offset.

For a PIMMS-like synchrotron and mixed \hetwo and \csix{} beams with ion fraction $r=0.2$, preliminary estimates indicate a range shift of  ${< \SI{0.1}{\milli\meter}}$ across the entire clinical energy range. At $r= 0.4$, for example, the  range shift would reach ${< \SI{0.2}{\milli\meter}}$.
These values confirm that range shifts for expected clinical mixing ratios ($r=0.1$ to $0.2$) fall well within reported clinical range tolerances, e.g. $0.3$\,mm in refs.~\cite{pivi_status_2019,grevillot:commissioning}. Nevertheless, further quantification of this effect, experimental validation, and assessment of its implications for energy verification systems~\cite{moser2012energy} are recommended for future studies.


\section{Conclusion}\label{sec:conclusion}

We have demonstrated a practical, noninvasive method to determine the ion composition of
dual-species beams with relative mass-to-charge offsets of order $10^{-4}$ in low- and medium-energy
ion synchrotrons with radial RF regulation loops.
An analytical model was derived that relates RF frequency offsets to ion fractions. The method was validated
with \textsc{Xsuite} tracking simulations and demonstrated experimentally at MedAustron, where the
RF-based mixing ratio estimates were found to be in good agreement with independent ion composition estimates of the extracted beam. 

The method is now used at MedAustron for mixed-beam commissioning and diagnostics during user experiments. It offers potential for developing a mixing ratio interlock system, with further improvements expected under optimized measurement conditions.

\section{Acknowledgments}
 The financial support of the Austrian Ministry of Women, Science, and Research is gratefully acknowledged for providing beam time and research infrastructure at MedAustron. Part of the research underlying this work was funded as part of the RTI-Strategy Lower Austria 2027 (FTI23-D-031). 
 The authors warmly thank C. Schmitzer (MedAustron) for collaboration on the mixed-beam project as well as the MedAustron OPS and TBU departments for their collaboration and support with measurements at the MedAustron synchrotron, including L. Adler, N. Gambino, T. Gere, T. Margreiter, F. Plassard, D. Prokopovich, V. Rizzoglio,  M. Sakelsek, and M. Wolf. 
 We are further grateful to P. Wurzer (MedAustron) for assistance with \cref{fig:schematicMixedBeamIrrad}, as well as F. Ulrich-Pur (TU Wien), M. Pullia (CNAO), A. Pastushenko (GSI) and the TU Wien Accelerator Physics students for discussion on related mixed-beam delivery and detection concepts.
 
\vspace{2mm}
E. R.: conceptualization, methodology, investigation, formal analysis, writing--original draft. M. K.: methodology, investigation, writing--review and editing. D. O.: methodology, writing--review and editing. H. F.: methodology. K. H.: investigation, writing--review and editing. N. O.: investigation, formal analysis, writing—review and editing.

\appendix

\section{\texorpdfstring{Validation with Xsuite simulations}{Appendix}}\label{sec:app:benchmark}
\Cref{eq:dff_for_r_new,eq:r_for_dff_new} are validated using the the beam dynamics simulation framework \textsc{Xsuite}~\cite{iadarola_xsuite_2024}, which allows for tracking of particles with non-nominal $m/q$. 
The simulations employed \texttt{xtrack}~v0.89.4 and are used to numerically determine the revolution frequency offset
$\Delta f$, which is required to regulate the charge-weighted radial displacement
${\langle x \rangle_\mathrm{q,mix}}$ to a specified ${x_\mathrm{RL}}$ for mixed \hetwo{} and \csix{} beams with ion fractions ${r=\qtyrange{0}{1}{}}$.
The simulations are prepared by initializing a generic PIMMS lattice with an on-momentum \csix{} as reference particle and $E_\mathrm{kin} = \SI{262.3}{\mega\electronvolt\per\u}$ as reference energy. This reference particle moves on a closed orbit with length $C$, and features the revolution frequency $f$ and the relativistic factors $\beta$ and $\gamma$.

As a substitute for an explicit radial feedback loop, the required RF frequency offset $\Delta f$ is determined using the Nelder-Mead optimization algorithm~\cite{NelderMead}, as implemented in \textsc{SciPy}~\cite{2020SciPy-NMeth}.
The algorithm varies $\Delta f$ to minimize the cost function
\begin{equation}\label{eq:costfunction}
  \mathcal{C} =
  \left|
    \left\langle x \right\rangle_{\mathrm{q,mix}} - x_\mathrm{RL}
  \right|.
\end{equation}
Each iteration of the optimization proceeds as follows.
\begin{enumerate}[label=(\arabic*)]
    \item \textit{Application of an RF frequency shift:}
    In \textsc{Xsuite}, applying a shifted frequency ${f_\mathrm{RF} = f + \Delta f}$ not only implies that this value is assigned as new frequency to the
    \texttt{xtrack.Cavity} element, but also that a corresponding  shift in the longitudinal coordinate $\zeta$,
    \begin{equation}
        \Delta \zeta = C \, \dfrac{\Delta f}{f},
    \end{equation}
    is assigned via the \texttt{xtrack.ZetaShift} element.
    The latter is the standard method for programming off-momentum operation in \textsc{Xsuite}~\cite{xtrack_offmomentum_doc}. 
    
    \item \textit{Initialization of a} \csix{} \textit{test particle close to the synchronous}
    \csix{} \textit{particle:}
    The test particle is initialized with the longitudinal phase space coordinates
    $\left(\zeta_\mathrm{C}, \delta_\mathrm{C} \right) = \left(0,\frac{\Delta f / f}{\eta}\right)$, with $\eta$ being the linear slippage factor. Due to numerical and linearization errors, the test particle is initialized close to, but not exactly on, the phase space coordinates of the synchronous particle and therefore performs small synchrotron oscillations around it.
    Since the test particle will be tracked over multiple synchrotron periods in step~(5),
    these small initial offsets will not affect the averaged solution.
    
    \item \textit{Initialization of a} \hetwo{} \textit{test particle close to the synchronous}
    \hetwo{} \textit{particle:} 
    The initial momentum deviation of \csix{}, $\delta_\mathrm{C}$, is then used to compute $\gamma_\mathrm{C}$, $\beta_\mathrm{C}$ and eventually the offset ${\delta}_\mathrm{sp}$ between \hetwo{} and \csix{} with \cref{eq:hat_delta}. Subsequently $\beta_\mathrm{He}$, obtained via \cref{eq:beta_expansion}, is used to determine the momentum deviation $\delta_\mathrm{He}$ of \hetwo{} with respect to the lattice reference particle. The \hetwo{} test particle is initialized with the longitudinal phase space coordinates
    $\left(\zeta_\mathrm{He}, \delta_\mathrm{He} \right) = \left(0, \delta_\mathrm{He}\right)$.

    \item \textit{Tracking and estimation of radial loop offset:}
    Both test particles are tracked over multiple synchrotron periods. The average radial displacements at the radial loop pick-up, $\langle x \rangle_\mathrm{C}$ and $\langle x \rangle_\mathrm{He}$, are obtained by averaging the respective offsets over multiple synchrotron periods, thereby averaging out radial oscillations arising from small-amplitude synchrotron motion. For a given ion fraction $r$, $\langle x \rangle_{\mathrm{q,mix}}$ is subsequently computed using \cref{eq:avg_x_q_mix}. 

    \item \textit{Computation of cost function:} The cost function $\mathcal{C}$, \cref{eq:costfunction}, is evaluated and returned to the optimization algorithm, which iteratively updates $\Delta f$ until convergence is achieved.
\end{enumerate}

\bibliography{references_ratiopaper.bib}

\end{document}